\documentclass[conference]{IEEEtran}
\usepackage{cite}
\usepackage{amsmath}
\usepackage[linesnumbered,ruled]{algorithm2e}
\usepackage{commath}
\usepackage{amsfonts}
\usepackage{graphicx}
\usepackage{subcaption}
\usepackage{float}
\graphicspath{ {images/} }
\usepackage{fancyvrb}
\usepackage{url}

\usepackage{amssymb}
\usepackage{amsthm}
\usepackage{ragged2e}
\usepackage{array,multirow}
\usepackage{textcomp}

\usepackage{eso-pic}
\newcommand\AtPageUpperMyright[1]{\AtPageUpperLeft{%
 \put(\LenToUnit{0.1\paperwidth},\LenToUnit{-1cm}){%
     \parbox{1.0\textwidth}{\raggedleft\fontsize{9}{11}\selectfont #1}}%
 }}%
\newcommand{\conf}[1]{%
\AddToShipoutPictureBG*{%
\AtPageUpperMyright{#1}
}
}

\newcommand{\placetextbox}[3]{
 \setbox0=\hbox{#3}
 \AddToShipoutPictureFG*{ \put(\LenToUnit{0.23\paperwidth},\LenToUnit{#2\paperheight}){\vtop{{\null}\makebox[0pt][c]{#3}}}
 }
 }
 \placetextbox{.23}{0.08}{978-1-7281-6309-3/19/\$31.00~
\copyright2019 IEEE \hfill}

\title{Comparisonal study of Deep Learning approaches on Retinal OCT Image}

\author{\IEEEauthorblockN{Nowshin Tasnim, Mahmudul Hasan, Ishrak Islam}
\IEEEauthorblockA{Department of Computer Science and Engineering\\
Khulna University of Engineering and Technology, Bangladesh\\
Email: nowshin1997@gmail.com, mahmudul@cse.kuet.ac.bd, ishrak.islam@gmail.com}
}

\begin{document}
\maketitle
\conf{\textbf{International Conference on Innovation in Engineering and Technology (ICIET) 23-24 December, 2019}}

\begin{abstract}
In medical science, the use of computer science in disease detection and diagnosis is gaining popularity. Previously, the detection of disease used to take a significant amount of time and was less reliable. Machine learning (ML) techniques employed in recent biomedical researches are making revolutionary changes by gaining higher accuracy with more concise timing. At present, it is even possible to automatically detect diseases from the scanned images with the help of ML. In this research, we have taken such an attempt to detect retinal diseases from optical coherence tomography (OCT) X-ray images. Here, we propose a deep learning (DL) based approach in detecting retinal diseases from OCT images which can identify three conditions of the retina. Four different models used in this approach are compared with each other. On the test set, the detection accuracy is 98.00\% for a vanilla convolutional neural network (CNN) model, 99.07\% for Xception model, 97.00\% for ResNet50 model, and 99.17\% for MobileNetV2 model. The MobileNetV2 model acquires the highest accuracy, and the closest to the highest is the Xception model. The proposed approach has a potential impact on creating a tool for automatically detecting retinal diseases.
\end{abstract}
\begin{IEEEkeywords}
Disease Detection, Deep Learning, Image Analysis
\end{IEEEkeywords}

\section{INTRODUCTION}
Retinal diseases are the most common cause of losing eye-sight at an early age. Most of them cause visual symptoms in the retina, a thin layer of tissue which is situated on the inside back wall of one's eye. Some of the retinal diseases are- choroidal neovascularization (CNV), drusen, and diabetic macular edema (DME) which are shown in Fig. \ref{fig: (a) (b) (c) (d)}. These diseases can be detected from the retinal OCT scanned image. In Normal OCT scan, the choroid layer has no hollow space, no damage in the fovea, or extra material between membrane and pigment. Retinal diseases can affect the choroid layer causing an effect on eye-sight severely. The layer can be teared up or bloated in some places or hole can be created in some areas. Myopia, vision loss, macular degeneration, etc., can occur because of the affected layer. Hence, this article's primary focus lies in the detection of retinal diseases from OCT scanned images.
\par
In the case of traditional approaches, automatic detection of retinal diseases includes a preprocessing unit with image quantization, downsampling, and segmentation. After preprocessing, these images are then used to train a shallow neural network which takes a considerable amount of time as a huge amount of data is required \cite{goldbaum1996automated}. Because of the limitations of the shallow network, nowadays transfer learning techniques (one of the DL techniques) have become more widespread \cite{liu2011automated}, \cite{krizhevsky2012imagenet}.
\par
In this work, four CNN models- Vanilla CNN, MobileNetV2, ResNet50, and Xception network are employed to detect the category of diseases from retinal OCT scanned images. The images are mainly of 4 categories and have variable sizes. By preprocessing, each image is transformed into a definite shape. These transformed images are passed to the model for training. After training, these models are evaluated on a separated test data to measure performance. The performance metrics of each model are depicted in Fig. \ref{Validation accuracies}. No extra image processing based feature extraction methods is adopted in this work. For this reason, it can work with any data. The primary goal of this work is to help the patient and doctor and make the diagnosis automated and faster. Another goal is to enhance the performance of the analysis by increasing accuracy.
\par
The specific contributions of our work are as follows:
\begin{itemize}
    \item We use images of variable sizes for training the models. To the best of our knowledge, for the first time such kind of work is done to handle variable sizes of OCT images to train the models.
    \item We train the models using batches of data from different categories at the same time to maintain randomness (including around 44.57\% CNV, 13.59\% DME, 10.32\% Drusen, and 31.52\% Normal data among 84,452 images). 
    \item We demonstrate different robust models of deep convolutional neural network for detection of diseases and demonstrate a comparison of performance measure among them. 
\end{itemize}
\par
The remaining section consists of the following:
Some related works are discussed in section \ref{Sec: Related work}. In section \ref{Sec: Methodology}, the architecture for this work, description of the dataset, and evaluation criteria are given. Experimental analysis and results are briefly discussed in section \ref{Sec: Exp and result analysis}. At last, the paper is concluded in section \ref{Sec: Conclusion}.

\section{RELATED WORKS}
\label{Sec: Related work}
In \cite{srinivasan2014fully}, Srinivasan et al. propose a fully automated way using sparsity-based block-matching and 3D-filtering (BM3D) to remove noise for detecting retinal diseases like- DME and age-related macular degeneration (AMD) from the spectral domain (SD) OCT volume images. In this method, the fraction of volume correctly classified for AMD and DME classes are 100\%, which is the indication of overfitting problem. However, in the case of a regular class, it is 86.67\%.

Burlina et al. describe a technique in which they used a pre-trained DCNN to detect AMD in \cite{burlina2016detection}. This approach has achieved quite an excellent preliminary result.

By using B-scans of OCT, a fully automated method to detect lesion activity of AMD is studied by Chakravarthy et al. in \cite{chakravarthy2016automated}. They have tested retinal specialist (RS) grading versus Notal OCT analyzer (NOA) for faster treatment purpose and got an accuracy of 91\%.

CNV, DME, Drusen, and their medical diagnoses are identified using CNNs and transfer learning by Kermany et al. in \cite{kermany2018identifying}. Collected images quality is reviewed initially and then is fed into the AI system. They have achieved accuracy of 96.6\% in a multi-class comparison.

In \cite{vahadane2018detection}, two-step framework to detect DME in OCT scans using patched based deep learning (DL) is proposed by Vahadane et al. They have used Dijkstra's shortest path algorithm for detecting candidate patches for fluid-filled regions, Otsu thresholding technique for extraction. For hard regions, local peak detection is used. After these, they have used a DCNN to predict the label of those patches. The accuracy of this approach is better than using a frame-level DCNN classifier, which is proven by them.

Using artificial intelligence approach, fully automated way to detect and quantify of macular fluid - Schlegl et al. describe intraretinal cystoid fluid (IRC) and subretinal fluid (SRF) in macular OCT scan which causes AMD and DME in \cite{schlegl2018fully}. DCNN is used for mapping images to corresponding labels, and high accuracy is achieved in that process.

In \cite{rong2019surrogate}, Rong et al. have used CNNs and suggested a classification method of surrogate-assisted to classify retinal OCT images automatically. They have reduced the noise from the image and applied thresholding and morphological dilation for extraction purpose. From there, they have generated surrogate images which are used for training. The accuracy is 97.83\% for their local database, and  98.56\% for the public database.

Li et al. have used the DL method VGG-16 network in \cite{li2019fully} for detecting retinal disorders in a fully automated way. Image normalization is used to label the images, and image denoising is avoided to get more accuracy for original data. Using this process, they have achieved prediction accuracy up to 98.6\%.

By using VGG-19 model, automated AMD detection combining OCT and fundus images are shown by Yoo et al. in \cite{yoo2019possibility} where five different models are used. An accuracy of 82.6\% and 83.5\% are achieved from only OCT image-based DL model and only fundus image based DL model whereas, in case of a combined image model, the accuracy is 90.5\%.

\section{Methodology}
\label{Sec: Methodology}
\begin{figure}[h]
  \centering
  \includegraphics[scale=.4]{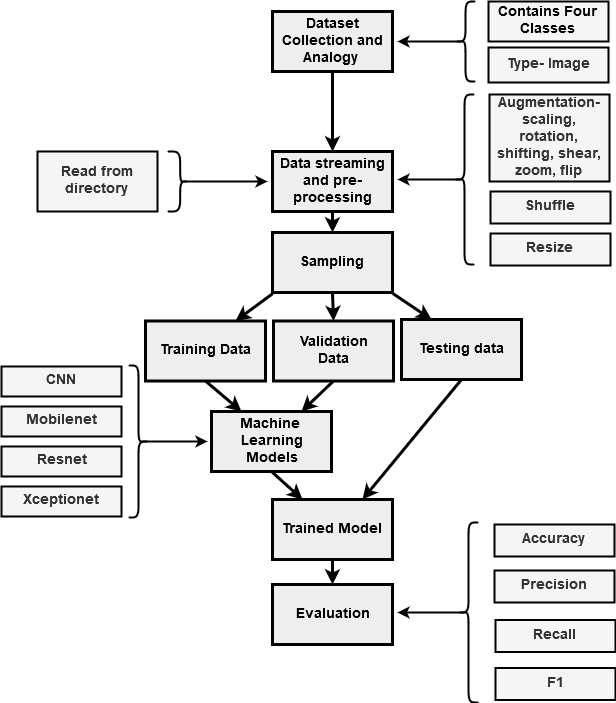}
  \caption{Working Procedure}
  \label{fig: working procedure}
\end{figure}

The whole architecture for the analysis of OCT image data requires several phases which are described in the following. Fig. \ref{fig: working procedure} illustrates the overall workflow of the research.

At the first stage of the workflow, raw image data goes through streaming and preprocessing pipeline. If a large amount of data are processed together for training, it causes overload on RAM, which may result in a system crash because of memory limitations. However, in this case, we have used data streaming, which takes data from the directory as batches. Our streamer can automatically detect the folders of classes for training and testing from the given directory for reading files. So, less code is required for reading a file than other cases. While streaming of the data from the directories, some filtering steps are executed dynamically at runtime. These filtering processes are data augmentation, data shuffling, and resizing of data. Augmentation step creates synthetic images by varying positioning, rotating, flipping, and shearing the real image data. After augmentation, we have resized both synthetic and original image into a fixed size to make every image uniform. On this uniform-sized images, we have applied the normalization process. At the final stage of data preprocessing pipeline, we have shuffled and partitioned the whole resulted images into training and testing sets. Shuffling is necessary for removing biases towards one particular class. The ratio of training, validation, and testing data split is 98.816:0.038:1.146. We have trained DL and transfer learning techniques vanilla CNN, MobileNetV2, ResNet50, and XCeption network with the training data in the training model phase of the workflow. After each epoch of training, we get the training accuracy curve of a model. At the same time, the validation accuracy curve of that model is also generated using validation dataset. At the last stage of the workflow, we have used different evaluation metrics and test data to evaluate the models that have been trained on the training data after 15 epochs of each model, which is known as the final model. Evaluation metrics consist of a confusion matrix, accuracy, precision, sensitivity, and f1 score for our proposed work.

\subsection{Dataset Collection and Description}
\begin{figure}[h]
  \centering
  \includegraphics[scale=.13]{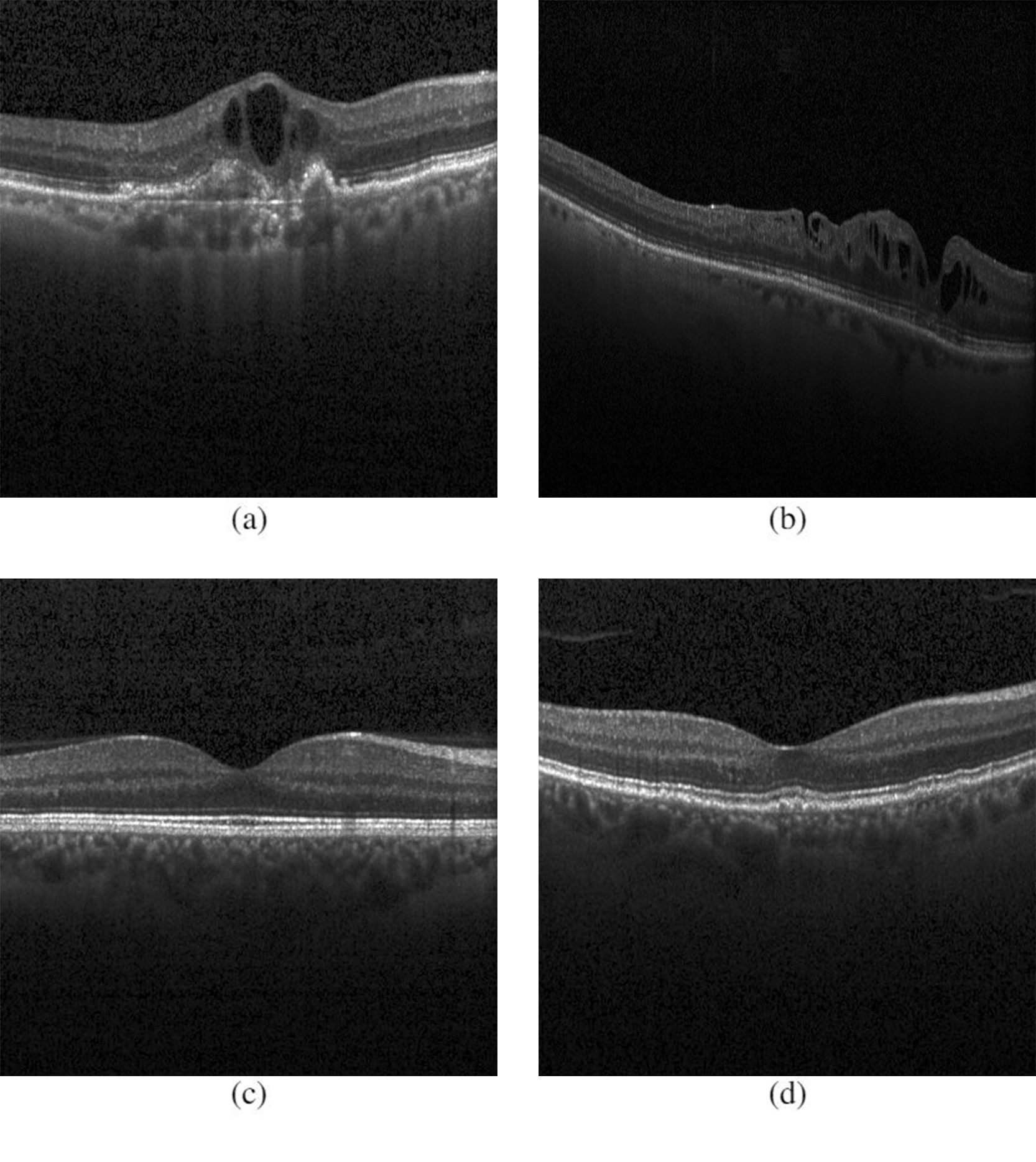}
  \caption{(a) CNV (b) DME (c) DRUSEN (d) NORMAL}
  \label{fig: (a) (b) (c) (d)}
\end{figure}

We have collected the dataset from popular website Kaggle\footnote{https://www.kaggle.com/paultimothymooney/kermany2018}. The whole dataset is organized into three folders (train, test, validation) containing sub-folders for four image categories (NORMAL, CNV, DME, DRUSEN) having total 84,484 X-Ray images (JPEG) with various shapes. Among these images for training and validation purpose, 83,484  and 32 images are used. We set aside 968 images for the test set. In Fig. \ref{fig: (a) (b) (c) (d)}, one figure from each category is given. From the figures, we can see the normal patient's eye in Fig. \ref{fig: (a) (b) (c) (d)}(d) and rest of the figures are of the infected eye.

New blood vessels which are created in the choroid layer of the eye is known as Choroidal neovascularization (CNV). The blood vessels are created from a break in the Bruch membrane. It is the main cause of losing central vision. Central Vision is the process in which light changes into nerve signals using millions of cells that inform the brain what the person is seeing. CNV is associated with extreme myopia, malignant myopic degeneration, and age-related developments. The break in the membrane can be seen as a hollow space in the retinal OCT x-ray in Fig. \ref{fig: (a) (b) (c) (d)}(a).

Diabetic macular edema (DME) is a complication of diabetes. Fluid accumulation in the retina's macula due to leaking blood vessels is the cause of DME. This disease makes the macula swell and affects the center of the macula, known as the fovea. In the OCT scanned image given in Fig. \ref{fig: (a) (b) (c) (d)}(b), the defect in fovea can be seen as small holes in the membrane. It causes blurred vision, wavy vision, double vision, a sudden change in eye floaters, and faulty perception of colors. It is also associated with low levels of protein in body fluids, high blood pressure, retention of fluid, and high levels of fats in the blood.

Between Bruch's membrane and the retinal pigment epithelium of the eye, if tiny yellow or white accumulations of extracellular material are build up, it is known as Drusen. They are like tiny pebbles of debris. Drusen is associated with aging and macular degeneration which causes central vision loss.

The dataset distribution among these four categories is given in the Table \ref{Frequency Distribution of Considered Attacks}.

\begin{table}[h]
    \centering
    \caption{Frequency Distribution of Retinal OCT Image}
    {\setlength{\extrarowheight}{1.5pt}%
    \begin{tabular}{|c|c|c|c|c|}
    \hline
    \textbf{Data}&\textbf{Training set}&\textbf{Testing Set}&\textbf{\% of}&\textbf{\% of}\\
    &\textbf{Frequency}&\textbf{Frequency}&\textbf{Training}&\textbf{Testing}\\ \hline
    CNV&37205&242&44.57&25\\ \hline
    DME&11348&242&13.59&25 \\ \hline
    DRUSEN&8616&242&10.32&25 \\ \hline
    NORMAL&26315&242&31.52&25 \\ \hline
    Total&83484&968&100&100 \\ \hline
    \end{tabular}}
    \label{Frequency Distribution of Considered Attacks}
\end{table}

\subsection{Deep Learning Models}
DL approaches are so far, the most stable strategy to handle image type data. Alongside the convolutional neural network, models like XCeption, MobileNetV2, ResNet have been employed in this work.
\subsubsection{Convolutional Neural Network (CNN)}
In the Convolutional Neural Network (CNN), each image data is flattened into vectors without losing features. As a result, the number of parameters involved and reusability of weights is reduced, and functional prediction result is obtained. The kernel, having the same depth as an image for each channel, is used to maintain valid and same padding and is shared by all the nodes of a layer. So, the resultant output has fewer nodes than a vanilla neural network. To reduce the spatial size, Pooling layer is also used, which can be of two types- max pooling and average pooling. Between these two pooling, max-pooling is better because it returns the maximum value from the part of the image covered by the kernel and performs as a noise suppressant. This algorithm is used in various architecture like- ResNet, Xception, VGGNet, and other DL techniques.

The vanilla CNN employed for this work developed with one input layer, eleven hidden layers, and one output layer. The model architecture for CNN is as following: 
\begin{itemize}
    \item Input Layer (Shape = 150x150x3)
    \item Convolutional Layer (Filter Size =  3x3, \# of filters = 64, activation = Relu)
    \item Max Pooling Layer (Filter size = 2x2)
    \item Convolutional Layer (Filter Size =  3x3, \# of filters = 64, activation = Relu)
    \item Max Pooling Layer (Filter size = 2x2)
    \item Convolutional Layer (Filter Size =  3x3, \# of filters = 128, activation = Relu)
    \item Max Pooling Layer (Filter size = 2x2)
    \item Convolutional Layer (Filter Size =  3x3, \# of filters = 128, activation = Relu)
    \item Max Pooling Layer (Filter size = 2x2)
    \item Flatten Layer
    \item Dropout (Probability = .5)
    \item Dense Layer (Shape = 512, activation = Relu)
    \item Dense Layer (Shape = 4, activation = softmax)
\end{itemize}
\subsubsection{Xception Net}
The Xception stands for 'Extreme Inception' which uses a modified depth wise separable convolution\cite{chollet2017xception}. This modification is to have a pointwise convolution followed by a depth wise convolution instead of having the usual another way around. There are also residual connections added among the depth wise separable convolution layers. Finally, it takes inception like architecture with three flow networks added sequentially. The three flows are - entry flow, middle flow, and exit flow. In this work, following layers are embedded with the base architecture of Xception: Input Layer (Shape = 150x150x3), Flatten layer, Dense layer (Shape = 1024, activation = Relu), Dropout (Probability = .2), Dense layer (Shape = 4, activation = softmax).

\subsubsection{ResNet50}
In Residual network (ResNet), the layers are constructed as identity mapping, each taking the output of the previous layer as the input using shortcut connection. So, no extra parameter is needed. Between 18-layer and 34-layer net, 34-layer net has less error and better accuracy \cite{he2016deep}. ResNet50 is Residual network with 50 layers where every two layers in the 34-layer net are replaced with 3-layer bottleneck block. For this work, the model architecture of ResNet50 is embedded with following layers: Input Layer (Shape = 150x150x3), Flatten layer, Dense layer (Shape = 1024, activation = Relu), Dropout (Probability = .2), Dense Layer (Shape = 4, activation = softmax).

\subsubsection{MobileNetV2}
MobileNetV2 uses inverted residual structure in which nonlinearities of thin layers are removed \cite{sandler2018mobilenetv2}. It works with two types of blocks. The basic building block is 19 bottleneck depth-separable convolution layers with residuals with stride 1 in one block and stride 2 in another block, which is preceded by fully convolutional layer with 32 filters. The first layer is 1x1 convolutional with Relu6, the second layer is depth wise convolution, and the third layer is 1x1 convolutional without nonlinearities using kernel size 3x3 for each layer. In the case of object detection, higher accuracy can be gained by this model. For this work, following layers are embedded with the MobileNetV2: Input layer (Shape = 150x150x3), Flatten layer, Dense layer (Shape = 1024, activation = Relu), Dropout (Probability = .2), Dense layer (Shape = 4, activation = softmax).

\subsection{Evaluation Criteria}
\subsubsection{Confusion Matrix}
The Confusion matrix, also known as the error matrix, can be used for visualization of a model's performance as the summary of prediction result for each class is accounted here \cite{hasan2019attack}. We can easily find the confusion or errors made by a classifier by checking the confusion matrix table. From this table, we can get the value of True Positive (TP), True Negative (TN), False Positive (FP), and False Negative (FN). Let us assume, Cl$_i$ denotes a class out of four classes. The definition of TP, TN, FP, FN for Cl$_i$ are as follows:
\begin{itemize}
    \item TP(Cl$_i$) = All the Cl$_i$ instances that are predicted correctly as Cl$_i$.
    \item TN(Cl$_i$) = All the instances of non Cl$_i$ that are not predicted as Cl$_i$.
    \item FP(Cl$_i$) = All the instances of non Cl$_i$ that are predicted as Cl$_i$.
    \item FN(Cl$_i$) = All the Cl$_i$ instances that are not predicted as Cl$_i$.
\end{itemize}

\subsubsection{Accuracy}
The measurement of how accurately a classifier classifies a data is known as accuracy \cite{hasan2019sentiment}. The equation for calculating accuracy is as follows:
\begin{equation}
    \text{Accuracy } = \frac{\text{T.P.}+\text{T.N.}}{\text{T.P.}+\text{T.N.}+\text{F.P.}+\text{F.N.} }
    \label{accuracy equation}
\end{equation}
Where, T.P. = True Positive, T.N. = True Negative, F.P. = False Positive and F.N. = False Negative.

\subsubsection{Precision}
Precision refers to the ratio of total correct positive results and total predicted positive results by the classifier. The equation of sensitivity is the following:
\begin{equation}
    \text{Precision} =
    \frac{\text{T.P.}}{\text{T.P.}+\text{F.P.} }
    \label{Precision equation}
\end{equation}

\subsubsection{Sensitivity}
Sensitivity is known as the true positive rate or recall, which refers to the proportion of correct positive data points respect to all positive data points. The equation of sensitivity is the following:
\begin{equation}
    \text{Sensitivity} =
    \frac{\text{T.P.}}{\text{T.P.}+\text{F.N.} }
    \label{sensitivity equation}
\end{equation}

\subsubsection{F1-Score}
In a model, the weighted average of Precision and Recall is known as F1 score \cite{hasan2019sentiment}. The
equation of the F1 score is the following:
\begin{equation}
    \text{F1 Score } = \frac{2*\text{T.P.}}{2*\text{T.P. }+\text{ F.P. }+\text{ F.N.} }
    \label{f1 equation}
\end{equation}

\section{Experimental Analysis and Result Analysis}
\label{Sec: Exp and result analysis}

\subsection{Experimental Setup}
Google Colab solely leverages this classification task. Colab consists of 1xTesla K80 (2496 CUDA cores), 1xsingle core hyper threaded Xeon Processors @2.3Ghz, 45MB Cache, 12.6 GB available RAM, and 320 GB available Disk Space.

\subsection{Result Analysis}
Table \ref{Confusion Matrix (Training and Testing)} describes the confusion matrix of our four models after training and testing. From the confusion matrix table, we found that Xception and MobileNetV2 can predict more accurately than ResNet50 and CNN. By using Eq. \ref{accuracy equation}, \ref{Precision equation}, \ref{sensitivity equation}, \ref{f1 equation}, we have calculated  the accuracy, precision, sensitivity and F1 Score respectively for each classifiers. The values are given in the Table \ref{Evaluation Metrics of Classifiers}. From Table \ref{Evaluation Metrics of Classifiers}, it is found that the training accuracy in all models is 89\% or above where the testing accuracy is 97\% or above. The training accuracy is higher in Xception, which is 93.90\%, and the testing accuracy is higher in MobileNetV2, which is 99.17\%. The precision in training is between 0.92 to 0.95, and in the case of testing, it is between 0.97 to 0.99, including the boundary values. For both cases, Xception has the highest precision, and  MobilenetV2 has the highest precision only in testing. The Sensitivity and F1 score in training are from 0.92 to 0.95, and in testing is from 0.97 to 0.99. Sensitivity and F1 score of 0.95 is achieved in Xception for training, and 0.99 is achieved for testing in Xception and MobileNetV2. From the four models, Xception and MobileNetV2 have yielded the best results. 

\begin{table}[h]
    \centering
    \caption{Confusion Matrix (Training and Testing)}
    {\setlength{\extrarowheight}{1.5pt}%
    \begin{tabular}{|c|c|c|c|c|c|c|}
    \hline
    \textbf{Model}&\textbf{}&\textbf{}&\textbf{CNV}&\textbf{DME}&\textbf{DRU}&\textbf{NRL}\\ \hline
    \parbox[t]{2mm}{\multirow{8}{*}{\rotatebox[origin=c]{90}{\textbf{CNN}}}}&
    \parbox[t]{2mm}{\multirow{4}{*}{\rotatebox[origin=c]{90}{\textbf{Training}}}}&CNV&35638&468&807&292\\ \cline{3-7}
    &&DME&527&9644&30&1147 \\ \cline{3-7}
    &&DRU&988&50&6229&1349 \\ \cline{3-7}
    &&NRL&109&219&48&25939 \\ \cline{3-7}
    \cline{2-7}
    &\parbox[t]{2mm}{\multirow{4}{*}{\rotatebox[origin=c]{90}{\textbf{Testing}}}}&CNV&240&2&0&0\\ \cline{3-7}
    &&DME&4&233&0&5 \\ \cline{3-7}
    &&DRU&0&0&242&0 \\ \cline{3-7}
    &&NRL&0&0&5&237 \\ \cline{3-7}
    \hline
    \parbox[t]{2mm}{\multirow{8}{*}{\rotatebox[origin=c]{90}{\textbf{Xception}}}}&
    \parbox[t]{2mm}{\multirow{4}{*}{\rotatebox[origin=c]{90}{\textbf{Training}}}}&CNV&35467&197&1391&150\\ \cline{3-7}
    &&DME&288&9813&59&1188 \\ \cline{3-7}
    &&DRU&272&11&7651&682 \\ \cline{3-7}
    &&NRL&36&25&145&26109 \\ \cline{3-7}
     \cline{2-7} 
    &\parbox[t]{2mm}{\multirow{4}{*}{\rotatebox[origin=c]{90}{\textbf{Testing}}}}&CNV&241&1&0&0\\ \cline{3-7}
    &&DME&3&238&0&1 \\ \cline{3-7}
    &&DRU&4&0&238&0 \\ \cline{3-7}
    &&NRL&0&0&0&242 \\ \cline{3-7}
    \hline
    \parbox[t]{2mm}{\multirow{8}{*}{\rotatebox[origin=c]{90}{\textbf{ResNet50}}}}&
    \parbox[t]{2mm}{\multirow{4}{*}{\rotatebox[origin=c]{90}{\textbf{Training}}}}&CNV&36093&251&776&85\\ \cline{3-7}
    &&DME&806&9184&43&1315 \\ \cline{3-7}
    &&DRU&1323&24&6244&1025 \\ \cline{3-7}
    &&NRL&354&99&209&25653 \\ \cline{3-7}
    \cline{2-7}
    &\parbox[t]{2mm}{\multirow{4}{*}{\rotatebox[origin=c]{90}{\textbf{Testing}}}}&CNV&241&1&0&0\\ \cline{3-7}
    &&DME&6&235&0&1 \\ \cline{3-7}
    &&DRU&18&0&222&2 \\ \cline{3-7}
    &&NRL&1&1&0&240\\ \cline{3-7}
    \hline
    \parbox[t]{2mm}{\multirow{8}{*}{\rotatebox[origin=c]{90}{\textbf{MobileNet-V-2}}}}&
    \parbox[t]{2mm}{\multirow{4}{*}{\rotatebox[origin=c]{90}{\textbf{Training}}}}&CNV&35399&182&1367&257\\ \cline{3-7}
    &&DME&427&10205&27&689 \\ \cline{3-7}
    &&DRU&389&27&7039&1161 \\ \cline{3-7}
    &&NRL&31&152&61&26071 \\ \cline{3-7}
    \cline{2-7}
    &\parbox[t]{2mm}{\multirow{4}{*}{\rotatebox[origin=c]{90}{\textbf{Testing}}}}&CNV&241&1&0&0\\ \cline{3-7}
    &&DME&2&240&0&0 \\ \cline{3-7}
    &&DRU&3&0&238&1 \\ \cline{3-7}
    &&NRL&0&0&1&241\\ \cline{3-7}
    \hline
    \end{tabular}}
    \label{Confusion Matrix (Training and Testing)}
\end{table}

\begin{table}[h]
    \centering
    \caption{Evaluation metrics of our study}
    {\setlength{\extrarowheight}{1.5pt}%
    \begin{tabular}{|c|c|c|c|c|c|}
    \hline
    \multicolumn{2}{|c|}{\textbf{Evaluation}} & \multicolumn{4}{c|}{\textbf{Classifiers}}\\
    \cline{3-6}
    \multicolumn{2}{|c|}{\textbf{Metrics}} & \textbf{CNN} & \textbf{Xception} & \textbf{ResNet-50} & \textbf{MobileNet-V-2}\\
    \hline
    \parbox[t]{2mm}{\multirow{4}{*}{\rotatebox[origin=c]{90}{\textbf{Training}}}}&\textbf{Accuracy} & 0.90& 0.9390 & 0.89 & 0.9388 \\\cline{2-6}
    &\textbf{Precision} & 0.93 & 0.95 & 0.92 & 0.94\\\cline{2-6}
    &\textbf{Sensitivity} & 0.93 & 0.95 & 0.92 & 0.94\\\cline{2-6}
    &\textbf{F1 Score} & 0.93& 0.95 & 0.92 & 0.94\\
    \hline
    \parbox[t]{2mm}{\multirow{4}{*}{\rotatebox[origin=c]{90}{\textbf{Testing}}}}&\textbf{Accuracy} & 0.98 & 0.9907 & 0.97 & 0.9917\\\cline{2-6}
    &\textbf{Precision} & 0.98 & 0.99 & 0.97 & 0.99\\\cline{2-6}
    &\textbf{Sensitivity} & 0.98 & 0.99 & 0.97 & 0.99\\\cline{2-6}
    &\textbf{F1 Score} & 0.98 & 0.99 & 0.97 & 0.99\\\cline{2-6}
    \hline
    \end{tabular}}
    \label{Evaluation Metrics of Classifiers}
\end{table}

\begin{figure}[h]
  \centering
  \includegraphics[scale=.5]{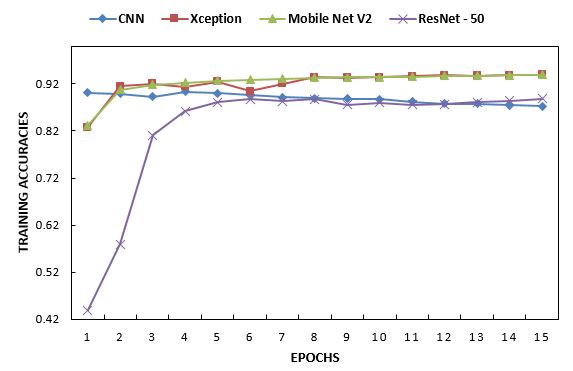}
  \caption{Training accuracy vs Epochs}
  \label{Training accuracies}
\end{figure}

For this work, we have used 15 epochs for training and validation purposes. The training accuracy vs. epochs graph is shown in Fig.  \ref{Training accuracies}. In the case of CNN, after first epoch the curve has decreased until the fourth epoch; then in the fourth epoch, it has increased and then decreased again. The Xception's curve has increased till the second epoch, then kept on decreasing and increasing until eighth epoch. After eighth epoch, it has become saturated. The MobileNetV2's curve has increased until the 2nd epoch. Then it has become saturated with higher accuracy like Xception. The Resnet50's curve has shown encouraging result until the fourth epoch as the accuracy has abruptly risen. Then the curve's changing has become slower and has reached around the saturation in sixth epoch with lower accuracy than Xception and MobileNetV2. Thus, the Xception and MobileNetV2 are more promising in case of training than the other two because of their higher accuracy.

\begin{figure}[h]
  \centering
  \includegraphics[scale=.5]{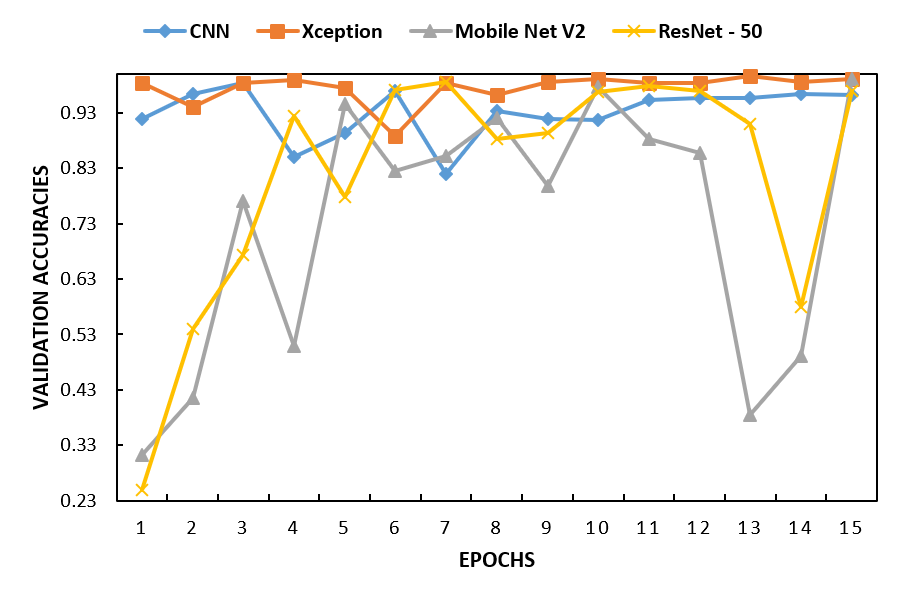}
  \caption{Validation accuracies vs Epochs}
  \label{Validation accuracies}
\end{figure}

The graph of validation accuracy for 15 epochs is shown in Fig. \ref{Validation accuracies}. The curve of vanilla CNN is upward and downward, and after a time, it has reached its saturation. The Xception's curve has reached the highest accuracy in most of the epochs. The MobileNetV2's curve has been changing abruptly upwards and downwards, and in the last epoch, it reached close to the Xception's accuracy. In the case of ResNet50's curve, drastic change of accuracy can be seen till the 4th epoch. After that, the accuracy has increased and decreased steadily, and in the last epoch, it has reached near CNN's accuracy. So, from Fig. \ref{Validation accuracies}, it can be said that Xception and MobileNetV2 have higher accuracy than the other two in testing.

From the Table \ref{tab: Quantitative result}, we can find that Kermany et al. has gained 96.60\% prediction accuracy with a sensitivity of 97.80\% in detection of retinal diseases using convolutional neural network \cite{kermany2018identifying}, where Li et al. has achieved a prediction accuracy of 98.60\% having sensitivity of 97.80\% for predicting retinal diseases in a fully automated way \cite{li2019fully}. In our method, to detect CNV, DME, and Drusen- these three diseases using the best model, accuracy of 99.17\% with a sensitivity of 99.00\% is achieved. These three works are on the same dataset. So, from the quantitative result, it is clear that our method is better among all the methods given in Table \ref{tab: Quantitative result} as it has the highest accuracy.

\begin{table}[h]
    \centering
    \caption{Proposed Comparative Analysis with Some Contemporary Methods}
    {\setlength{\extrarowheight}{1.5pt}%
    \begin{tabular}{|c|c|c|c|c|}
        \hline
            \textbf{Method} & \textbf{Accuracy} & \textbf{Sensitivity} & \textbf{Precision} & \textbf{F1-score} \\
        \hline
            Our method & 99.17\% & 99.00\% & 99.00\% & 0.99 \\
        \hline
            Kermany et al. \cite{kermany2018identifying} & 96.60\% & 97.80\% & N/A & N/A \\
        \hline
            Li et al. \cite{li2019fully} & 98.60\% & 97.80\% & N/A & N/A \\
        \hline
    \end{tabular}
    }
    \label{tab: Quantitative result}
\end{table}

\section{Conclusions}
\label{Sec: Conclusion}
In this experimental study, we demonstrate a better way to detect some retinal diseases from OCT X-ray and also show the comparison with other works \cite{kermany2018identifying},  \cite{li2019fully}. Our primary contribution is to work with variable sized images. We explored multiple deep learning models \cite{chollet2017xception}, \cite{he2016deep}, \cite{sandler2018mobilenetv2} to present the performances in terms of accuracy, sensitivity, precision, and f1 score \cite{hasan2019attack}, \cite{hasan2019sentiment}. We can claim that our approach can detect the discussed retinal diseases better as the accuracy is higher than state of the art. There is a limitation regarding the distribution of classes; that is, all classes are not equally distributed. If it can be overcome, the performance will increase. We have not used other image processing techniques. Maybe the use of image processing techniques could have led to an exciting result, which can make our work better and increase performance. We are leaving these issues for future works.
\bibliographystyle{unsrt}
\bibliography{ref}

\begin{thebibliography}{10}

\bibitem{goldbaum1996automated}
Michael Goldbaum, Saied Moezzi, Adam Taylor, Shankar Chatterjee, Jeff Boyd,
  Edward Hunter, and Ramesh Jain.
\newblock Automated diagnosis and image understanding with object extraction,
  object classification, and inferencing in retinal images.
\newblock In {\em Proceedings of 3rd IEEE International Conference on Image
  Processing}, volume~3, pages 695--698. IEEE, 1996.

\bibitem{liu2011automated}
Yu-Ying Liu, Mei Chen, Hiroshi Ishikawa, Gadi Wollstein, Joel~S Schuman, and
  James~M Rehg.
\newblock Automated macular pathology diagnosis in retinal oct images using
  multi-scale spatial pyramid and local binary patterns in texture and shape
  encoding.
\newblock {\em Medical image analysis}, 15(5):748--759, 2011.

\bibitem{krizhevsky2012imagenet}
Alex Krizhevsky, Ilya Sutskever, and Geoffrey~E Hinton.
\newblock Imagenet classification with deep convolutional neural networks.
\newblock In {\em Advances in neural information processing systems}, pages
  1097--1105, 2012.

\bibitem{srinivasan2014fully}
Pratul~P Srinivasan, Leo~A Kim, Priyatham~S Mettu, Scott~W Cousins, Grant~M
  Comer, Joseph~A Izatt, and Sina Farsiu.
\newblock Fully automated detection of diabetic macular edema and dry
  age-related macular degeneration from optical coherence tomography images.
\newblock {\em Biomedical optics express}, 5(10):3568--3577, 2014.

\bibitem{burlina2016detection}
Philippe Burlina, David~E Freund, Neil Joshi, Y~Wolfson, and Neil~M Bressler.
\newblock Detection of age-related macular degeneration via deep learning.
\newblock In {\em 2016 IEEE 13th International Symposium on Biomedical Imaging
  (ISBI)}, pages 184--188. IEEE, 2016.

\bibitem{chakravarthy2016automated}
Usha Chakravarthy, Dafna Goldenberg, Graham Young, Moshe Havilio, Omer Rafaeli,
  Gidi Benyamini, and Anat Loewenstein.
\newblock Automated identification of lesion activity in neovascular
  age-related macular degeneration.
\newblock {\em Ophthalmology}, 123(8):1731--1736, 2016.

\bibitem{kermany2018identifying}
Daniel~S Kermany, Michael Goldbaum, Wenjia Cai, Carolina~CS Valentim, Huiying
  Liang, Sally~L Baxter, Alex McKeown, Ge~Yang, Xiaokang Wu, Fangbing Yan,
  et~al.
\newblock Identifying medical diagnoses and treatable diseases by image-based
  deep learning.
\newblock {\em Cell}, 172(5):1122--1131, 2018.

\bibitem{vahadane2018detection}
Abhishek Vahadane, Ameya Joshi, Kiran Madan, and Tathagato~Rai Dastidar.
\newblock Detection of diabetic macular edema in optical coherence tomography
  scans using patch based deep learning.
\newblock In {\em 2018 IEEE 15th International Symposium on Biomedical Imaging
  (ISBI 2018)}, pages 1427--1430. IEEE, 2018.

\bibitem{schlegl2018fully}
Thomas Schlegl, Sebastian~M Waldstein, Hrvoje Bogunovic, Franz Endstra{\ss}er,
  Amir Sadeghipour, Ana-Maria Philip, Dominika Podkowinski, Bianca~S Gerendas,
  Georg Langs, and Ursula Schmidt-Erfurth.
\newblock Fully automated detection and quantification of macular fluid in oct
  using deep learning.
\newblock {\em Ophthalmology}, 125(4):549--558, 2018.

\bibitem{rong2019surrogate}
Yibiao Rong, Dehui Xiang, Weifang Zhu, Kai Yu, Fei Shi, Zhun Fan, and Xinjian
  Chen.
\newblock Surrogate-assisted retinal oct image classification based on
  convolutional neural networks.
\newblock {\em IEEE journal of biomedical and health informatics},
  23(1):253--263, 2019.

\bibitem{li2019fully}
Feng Li, Hua Chen, Zheng Liu, Xuedian Zhang, and Zhizheng Wu.
\newblock Fully automated detection of retinal disorders by image-based deep
  learning.
\newblock {\em Graefe's Archive for Clinical and Experimental Ophthalmology},
  257(3):495--505, 2019.

\bibitem{yoo2019possibility}
Tae~Keun Yoo, Joon~Yul Choi, Jeong~Gi Seo, Bhoopalan Ramasubramanian,
  Sundaramoorthy Selvaperumal, and Deok~Won Kim.
\newblock The possibility of the combination of oct and fundus images for
  improving the diagnostic accuracy of deep learning for age-related macular
  degeneration: a preliminary experiment.
\newblock {\em Medical \& Biological Engineering \& Computing}, 57(3):677--687,
  2019.

\bibitem{chollet2017xception}
Fran{\c{c}}ois Chollet.
\newblock Xception: Deep learning with depthwise separable convolutions.
\newblock In {\em Proceedings of the IEEE conference on computer vision and
  pattern recognition}, pages 1251--1258, 2017.

\bibitem{he2016deep}
Kaiming He, Xiangyu Zhang, Shaoqing Ren, and Jian Sun.
\newblock Deep residual learning for image recognition.
\newblock In {\em Proceedings of the IEEE conference on computer vision and
  pattern recognition}, pages 770--778, 2016.

\bibitem{sandler2018mobilenetv2}
Mark Sandler, Andrew Howard, Menglong Zhu, Andrey Zhmoginov, and Liang-Chieh
  Chen.
\newblock Mobilenetv2: Inverted residuals and linear bottlenecks.
\newblock In {\em Proceedings of the IEEE Conference on Computer Vision and
  Pattern Recognition}, pages 4510--4520, 2018.

\bibitem{hasan2019attack}
Mahmudul Hasan, Md~Milon Islam, Ishrak Islam, and MMA Hashem.
\newblock Attack and anomaly detection in iot sensors in iot sites using
  machine learning approaches.
\newblock {\em Internet of Things}, page 100059, 2019.

\bibitem{hasan2019sentiment}
Mahmudul Hasan, Ishrak Islam, and KM~Azharul Hasan.
\newblock Sentiment analysis using out of core learning.
\newblock In {\em 2019 International Conference on Electrical, Computer and
  Communication Engineering (ECCE)}, pages 1--6. IEEE, 2019.

\end{thebibliography}
 \end{document}